\documentclass[sigconf]{acmart}
\acmConference[EASE 2026]{The 30th International Conference on Evaluation and Assessment in Software Engineering}{9–12 June, 2026}{Glasgow, Scotland, United Kingdom}

\AtBeginDocument{%
  }

\usepackage{amsfonts}
\usepackage{algorithmic}
\usepackage{graphicx}
\usepackage{textcomp}
\usepackage{xcolor}
\usepackage{booktabs}
\usepackage{amsfonts}
\usepackage{listings}
\usepackage{todonotes}
\usepackage{subcaption}
\usepackage{tcolorbox}
\usepackage[utf8]{inputenc}
\usepackage{balance}

\usepackage{tcolorbox}

\newtcolorbox{answerbox}{
    colback=gray!10,      % Light gray background
    colframe=gray!60,     % Darker gray border
    boxrule=0.5pt,
    arc=0pt,              
    left=4pt, right=4pt, top=4pt, bottom=4pt,
    fontupper=\small
}

\lstdefinestyle{acmcode}{
    language=Java,
    basicstyle=\ttfamily\footnotesize,
    keywordstyle=\color{blue}\bfseries,
    commentstyle=\color{gray}\itshape,
    stringstyle=\color{red},
    breaklines=true,
    breakatwhitespace=true,
    frame=single, % Box around the code
    numbers=left,
    numberstyle=\tiny\color{gray},
    xleftmargin=10pt,
    columns=fullflexible,
    keepspaces=true
}

\begin{document}

\title[Sentence Transformer-based Filtering of Non-Actionable Alerts]{Towards Better Static Code Analysis Reports: Sentence Transformer-based Filtering of Non-Actionable Alerts}

\author{Tamás Aladics}
\email{aladics@inf.u-szeged.hu}
\orcid{0000-0002-4689-8878}
\affiliation{%
  \institution{University of Szeged}
  \institution{FrontEndART Ltd.}
  \city{Szeged}
  \country{Hungary}
}

\author{Norbert Vándor}
\email{vandor@inf.u-szeged.hu}
\orcid{0000-0001-8665-4360}
\affiliation{%
  \institution{University of Szeged}
  \city{Szeged}
  \country{Hungary}
}

\author{Rudolf Ferenc}
\email{ferenc@inf.u-szeged.hu}
\orcid{0000-0001-8897-7403}
\affiliation{%
  \institution{University of Szeged}
  \city{Szeged}
  \country{Hungary}
}

\author{Péter Hegedűs}
\email{hpeter@inf.u-szeged.hu}
\orcid{0000-0003-4592-6504}
\affiliation{%
  \institution{University of Szeged}
  \city{Szeged}
  \country{Hungary}
}

\renewcommand{\shortauthors}{Aladics et al.}

\begin{abstract}
Static code analysis (SCA) tools are widely used as effective ways to detect bugs and vulnerabilities in software systems. However, the reports generated by these tools often contain a large number of non-actionable findings, which can overwhelm developers to the point of ignoring them altogether -- this phenomenon is known as "alert fatigue". In this paper, we combat alert fatigue by proposing STAF: Sentence Transformer-based Actionability Filtering. Our approach leverages a transformer based architecture with sentence embeddings to classify findings into actionable and non-actionable categories. Evaluating STAF on a large dataset of reports from Java projects, we demonstrate that our method can effectively reduce the number of non-actionable findings while maintaining a high level of accuracy in identifying actionable issues. The results show that our approach can improve the usability of static analysis tools reaching an $\mathbf{F_{1}}$ score of 89\%, outperforming existing methods for SCA warning filtering by at least 11\% in a within-project setting and by at least 6\% in a cross-project setting. By providing a more focused and relevant set of findings, we aim to enhance the overall effectiveness of static analysis in software development. 
\end{abstract}



\maketitle

\section{Introduction}
As software systems grow, the complexity of their code-bases increases, the resource demands of development teams also rise, while security threats and exploits become more and more concerning. Many of the bugs and vulnerabilities in software systems are results of human error, such as typos, incorrect logic or failure to follow best practices. Answering these challenges, static code analysis (SCA) tools are widely used to automatically detect potential issues in code-bases. These tools analyze source code without executing it, usually by applying a set of predefined rules and patterns and are extensively used by developers in multiple stages of the software development life-cycle~\cite{Balachandran:2013,Bessey:2010}.

However, the reports generated by SCA tools often contain a large amount of findings, many of which are not relevant for the developers. Heckman et al. found that 35\% to 91\% of the warnings generated by SCA tools are false positives~\cite{Heckman:2011}. Moreover, even if a report is valid and indicates a genuine issue (i.e., a "true positive" warning), the developers may still choose to disregard it as it may not result in a defect (such as inappropriate variable naming) or it may not be relevant to the current context~\cite{Ayewah:2007,Katerina:2015}. Because of this, developers tend to be overwhelmed by the sheer volume of non-actionable findings, resulting in a phenomenon known as "alert fatigue", which leads to decreased productivity and sometimes even abandonment of the SCA tools altogether~\cite{Kharkar:2022,Baca:2010}.

To address these issues, various approaches have been explored, such as statistical probability approaches~\cite{Wei:2017, Kim:2007}, static program analysis for further filtering~\cite{Chen:2013, Muske:2015, Valdiviezo:2014}, and clustering techniques~\cite{Guo:2023}. In the recent years, machine learning (ML) techniques have also been applied to filter SCA reports as the ability of ML models to learn from vast datasets makes them well-suited for tackling the problem of identifying non-actionable warnings~\cite{Lee:2019,Kharkar:2022, Hegedűs:2022}. Additionally, the scalability of ML solutions enables potential integration into different sized projects. 

In this work, we propose a novel ML based approach that embeds the warning related artifacts (the warning message, warning line and source code context) to a high-dimensional vector space, aggregates them using a feature fusing component, then classifies them into the \verb|actionable| and \verb|non-actionable| categories. The overall model architecture (see Figure~\ref{fig:architecture}) follows human intuition, it uses a dedicated transformer-based component~\cite{Su:2024} to process the warning lines, which usually holds the most central information about the warning, while the warning message and surrounding lines (source code context) are embedded using a general purpose sentence transformer, mGTE~\cite{reimers-2020-multilingual-sentence-bert, Zhang:2024}.

We evaluated our model in both within-project and cross-project settings. In the within-project setting, our model achieves an $\text{F}_{1}$ score of 89\% and an Matthews Correlation Coefficient (MCC) of 0.87, outperforming existing approaches by at least 11\% and 4\%, respectively. While performance decreases when generalizing to unseen projects, STAF still leads in the cross-project setting, outperforming baselines by at least 16\% in $\text{F}_{1}$ score and 9\% in MCC. As the within-project setup is prone to randomness, we also conducted a 5-fold cross-validation which shows stable performance across all folds. Finally, our ablation study shows that the pre- and post-context embeddings are crucial, as removing both leads to a 10\% drop in $\text{F}_{1}$ score and an 11\% drop in MCC.

Our contributions can be summarized as follows:
\begin{itemize}
    \item We propose a novel approach to filter static analysis reports using a transformer-based architecture with sentence embeddings, which effectively reduces the number of non-actionable findings.
    \item We evaluate our approach on a large dataset of more than 1 million reports from Java projects, demonstrating its effectiveness in improving the usability of static analysis tools.
    \item We compare our solution with three baseline approaches -- DeepInfer\cite{Kharkar:2022}, ChatGPT\cite{Brown:2020} and PRISM\cite{Yang:2024}, showing its potential advantages in terms of key performance metrics such as F1 score and Matthews Correlation Coefficient (MCC).    
\end{itemize}

\section{Motivating example} \label{sec:motivation}
\begin{figure}[h]
    \centering
    \begin{subfigure}{\linewidth}
        \begin{lstlisting}[style=acmcode]
String localVarPath = "/fake/test-query-parameters";
...
List<Pair> localVarQueryParams = new ArrayList<Pair>();\end{lstlisting}
        \caption{Actionable warning: Explicit type arguments can be replaced by a diamond: `new ArrayList<>()`}
        \label{fig:actionable}
    \end{subfigure}
    
    \vspace{1em} % Space between subfigures

    \begin{subfigure}{\linewidth}
        \begin{lstlisting}[style=acmcode]
PARSER = new com.google.protobuf.AbstractParser<Metadata>() {
    @Override
    public Metadata parsePartialFrom(
        com.google.protobuf.CodedInputStream input,
        com.google.protobuf.ExtensionRegistryLite ext) {
        // Implementation details...
    }
};\end{lstlisting}
        \caption{Non-Actionable: Explicit type arguments can be replaced by a diamond: `new com.google.protobuf.AbstractParser<>()`
}
        \label{fig:non_actionable}
    \end{subfigure}

    \caption{Comparison of context-dependent warnings for Explicit Type Arguments. In (a), the refactoring is safe, in (b), the context of an anonymous class makes the warning non-actionable.}
    \label{fig:motivating_example}
\end{figure}

In this section, we present a motivational example to illustrate the nuances of warning actionability and its strong dependence on code context. This example is selected from the dataset used to train and evaluate the models discussed in this work~\cite{Koszo:2025}. While we have chosen a relatively simple ``code smell'' issue for clarity of explanation, the dataset contains significantly more complex instances where context plays an equally critical role in determining validity.

In Figure~\ref{fig:motivating_example}, we observe two code snippets triggering similar warnings with the same base message structure. Figure~\ref{fig:actionable} displays a warning from the \textit{openapi-generator}~\footnote{https://github.com/OpenAPITools/openapi-generator, accessed on 2025.11.28.} project with the message: ``\textit{Explicit type arguments can be replaced by a diamond: \texttt{new ArrayList<>()}}''. Conversely, Figure~\ref{fig:non_actionable} displays a warning from the \textit{nacos}~\footnote{https://github.com/alibaba/nacos, accessed on 2025.11.28.} project: ``\textit{Explicit type arguments can be replaced by a diamond: \texttt{new com.google.protobuf.AbstractParser<>()}}''.

Although both warnings suggest the same refactoring—replacing explicit type arguments with the diamond operator (\texttt{<>})—their actionability and readability implications differ significantly due to the context. In the actionable instance (Figure~\ref{fig:actionable}), the pre-context explicitly identifies the variable as a local query parameter connected to a list. Because this usage is already clear from the surrounding declaration lines, the application of the diamond operator does not necessarily make the code more readable compared to the explicit type, it is simply a syntactic cleanup. 

In contrast, in the non-actionable case (Figure~\ref{fig:non_actionable}), the use of the diamond operator is less desirable. Unlike the simple variable assignment in the actionable case, the explicit type here serves a clear documentation purpose within the complex structure of the anonymous class. More importantly, however, the syntactic environment—specifically the post-context indicating an anonymous class definition—renders the diamond operator invalid in many Java environments, as it can cause type inference failures during compilation.

Based on this brief example, it is apparent that the same warning type can be actionable or non-actionable depending entirely on the environment and code context. It is also evident that traditional SCA tools often struggle to resolve this issue effectively, potentially flagging lines where the suggested fix is syntactically invalid despite appearing aesthetically desirable. In our work, we aim to improve both the identification and resolution of these context-dependent warnings.

\section{Background and related works}
\subsection{Static Code Analysis}\label{sec:sca}
SCA tools are widely used to automatically and systematically analyze source code to detect potential issues, typically by matching patterns and rules against the codebase, without executing it. Because different programming languages have different syntax and semantic rules, SCA tools are usually language-specific. In our work, we focus on reports generated for Java projects, as it is one of the most widely used languages with a large ecosystem of SCA tools, such as SonarQube,\footnote{https://www.sonarqube.org, accessed on 2025.11.28.} SpotBugs,\footnote{https://github.com/spotbugs/spotbugs, accessed on 2025.11.28.} and PMD.\footnote{https://pmd.github.io/, accessed on 2025.11.28.} 
Also, the literature contains the highest number of publications related to filtering non-actionable Java alerts, making it a suitable choice for an extensive comparison and evaluation~\cite{Guo:2023}.

Though SCA tools have varying outputs, they generate reports that include at least two main components: the warning message and the line numbers corresponding to the alert. The warning message usually contains a description of the issue in natural language, while the line numbers indicate where the issue was found in the source code. In addition, developers often consider the source code context, which includes the lines of code surrounding the warning line, as it can provide additional information about the issue and its potential impact. We used these three components (warning message, warning line and source code context) as the input to our model.

\subsection{Filtering Non-Actionable Warnings}\label{sec:related-filtering}
As SCA tools became more prominent, the hindering problem of high rate of false-positive and non-actionable warnings became apparent too. To mitigate this, the research community proposed numerous approaches~\cite{Guo:2023}.

Statistical probability approaches consider metrics and measurements corresponding to the analyzed program. For example, Wei et al.~\cite{Wei:2017} used Android app user reviews to prioritize warnings by establishing links between them. Their method, called OASIS, considers these links' strength as basis of prioritizing the warnings. Kim et al.~\cite{Kim:2007} proposed to analyze warning fixes in software change history to infer statistical patterns between the time a warning is removed and the actionability of an SCA alert. They built upon the assumption that if a warning is shortly fixed after its discovery, it means it was urgent and valid, therefore, actionable while non-actionable alerts may stay in the code-base longer. In our work, we refrain from using such approaches as they require additional data that is not always available or might violate privacy policies, such as user reviews or software change history.

Another line of research investigates the extension of static code analysis with more complex techniques to further filter the alerts. Chen et al.~\cite{Chen:2013} utilized thread specialization to filter the false positives of data-race warnings, successfully reducing their number by ~89\%. Many researchers used model checking to eliminate false positive warnings by creating a mathematical model of a software component's behavior and checking it against the SCA tool's findings~\cite{Muske:2015, Valdiviezo:2014}. However, these approaches are often complex and require significant computational resources, making them less suitable for large code-bases or real-time analysis we aim to support.

Machine learning based approaches have become increasingly popular in recent years in software engineering in general and also in the case of SCA report filtering, leveraging similar techniques as statistical probability approaches but with being more data driven and, therefore, offering more flexibility and robustness. Koc et al.~\cite{Koc:2019} proposed using a Long Short-Term Memory (LSTM) recurrent neural network to filter false positives from a Java static analysis tool. Instead of using raw source code, they first summarized the program by computing a backward slice from the reported error line, which was then converted into a sequence of tokens and processed through a series of data preparation routines. Heckman and Williams \cite{Heckman:2009} evaluated 51 alert characteristics, including source code history and code churn, to identify actionable warnings. They concluded that optimal models and predictive features vary significantly across projects, necessitating project-specific training. In contrast to their approach, which depends on historical repository data, our STAF model leverages a transformer-based architecture to directly extract semantic meaning from the raw code context and warning messages. By utilizing deep contextual embeddings rather than manual feature engineering, STAF achieves strong, generalizable performance across different projects without relying on version control metadata.

Kharkar et al.~\cite{Kharkar:2022} proposed DeepInferEnhance, a supervised learning model, to reduce false positives from the Infer static analyzer~\cite{Calcagno:2011}. The model is a transformer based on the CodeBERTa architecture~\cite{Wolf:2020}, with a classifier head on top of it which was fine-tuned on a labeled dataset of Infer warnings. They also proposed a generative model, GTP-C\cite{Svyatkovskiy:2020} that uses the model's code completion recommendations: if GPT-C suggests adding a  null-check in the context of a potential null dereference, the warning is classified as legitimate.

Yang et al.~\cite{Yang:2024} investigated the effectiveness of various Code Representation Learning (CRL) techniques for reducing false positives in static bug detectors. They proposed an ensemble approach named PRISM, which aggregates the outputs of multiple top-per-forming CRL-based models to prioritize static analysis warnings. The PRISM method works by taking the results from several individual models — each combining a code representation (e.g., Word2Vec~\cite{Mikolov:2013}, FastText~\cite{Joulin:2017}) with a neural network (e.g., BiLSTM) — and using a voting strategy to determine the final classification for a warning.

We chose to compare our method against DeepInferEnhance and PRISM, as they are the most recent and relevant works in the field to the best of our knowledge. We omit the comparison with the work of Koc et al. as computing backward slices is not feasible for large codebases. GPT-C is also not considered, as it is specifically designed to find null dereference issues, which is a specific type of warning that is not representative of the general problem of non-actionable SCA warnings that we aim to address. Note that we also compare our results with the output of ChatGPT, see Section~\ref{sec:nlp}.

\subsection{Natural Language Processing}\label{sec:nlp}
With the introduction of transformer-based architectures, natural language processing (NLP) has seen significant advancements in recent years. Transformers, such as BERT~\cite{Devlin:2019} and its variants~\cite{Liu:2019,Feng:2020}, have revolutionized the field by enabling models to learn contextual representations of words and sentences. These models are pre-trained on large corpora and can be fine-tuned for specific tasks, such as text classification or named entity recognition.

In our work, we focus on source code as context which can be interpreted as long sentences, therefore employing high-quality embeddings is an essential area. However, standard BERT models are suboptimal for deriving high-quality sentence embeddings directly.  Reimers and Gurevych~\cite{Reimers:2019} pointed out that common approaches like averaging BERT's output vectors or using the [CLS] token's output result in poor sentence embeddings, often performing worse than older methods like averaged GloVe embeddings.  To address this quality issue, they proposed Sentence-BERT (SBERT), a modification of the pre-trained BERT model that uses a siamese network structure. This architecture is fine-tuned to create semantically meaningful sentence embeddings where similar sentences are mapped to similar vectors. Models based on this approach, known as Sentence Transformers, are specifically designed to produce superior sentence vector representations.

One such model is mGTE~\cite{Zhang:2024}, which improves multilingual text retrieval through a novel text encoder with Rotary Position Embedding (RoPE) and unpadding, pre-trained on an 8192-token context. mGTE's encoder outperforms same-sized state-of-the-art models, while its Text Representation Model (TRM) and reranker match or exceed the performance of larger models on long-context retrieval benchmarks, all while demonstrating higher training and inference efficiency. Since our objective was to provide a lightweight solution that can be run without requiring extensive computational resources, we chose mGTE as it is lightweight (305M parameters) while still being one of the best performing models on the MTEB benchmark's~\cite{Muennighoff:2022} coding and general text tasks.

Large language models (LLMs) are also heavily used for various NLP tasks, such as text generation, summarization and question answering. LLMs, such as GPT-3~\cite{Brown:2020} and its successors, have shown remarkable performance in generating human-like text and understanding complex language patterns. Note that these models are typically designed for text generation, rather than classification and to use them as such requires additional steps (e.g. prompt engineering). Still, since these models have been shown to perform well on various NLP tasks we also used ChatGPT, a state-of-the-art LLM, to compare our method against.

\section{Model Architecture}
\begin{figure*}[h]
  \centering 
  \includegraphics[width=\textwidth]{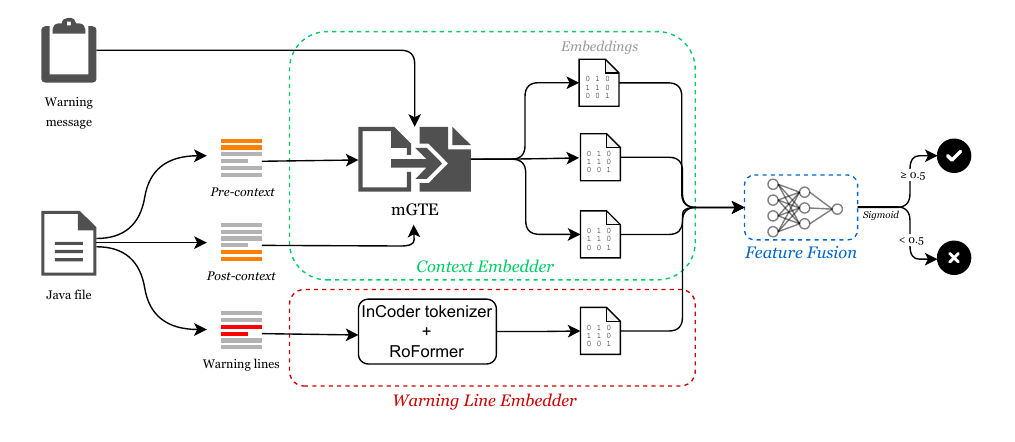}
  \caption{The overview of the model architecture for filtering non-actionable code alerts}
  \label{fig:architecture} 
\end{figure*}

In Figure~\ref{fig:architecture} we present the overall architecture of our model.   
The input to the model are the three core artifacts of static analysis reports, as outlined in Section~\ref{sec:sca}: the warning message, the warning line and the source code context. Developers usually follow human intuition, where the main focus is on the warning line, while the warning message and source code context are used to provide additional information about the issue. Our model architecture is designed to reflect this: the warning line is processed by a dedicated transformer-based component that is trained from scratch, while the warning message and source code context are processed by a general-purpose, pre-trained sentence transformer. In the following, we are going to describe the components of the model in more detail.

\subsection{Warning Line Embedder (WLE)}
The Warning Line Embedder (WLE, the red rectangle in Figure~\ref{fig:architecture}) is designed to process the warning lines as raw text input, parsed from the source code. 

\paragraph{Preprocessing} For some SCA tools it is possible to forcibly ignore a specific warning by adding a single-line comment to the warning line (e.g. ``\verb|// NOSONAR|'' in SonarQube\cite{sonarqube}), signaling non-actionability explicitly. Such comments would make the model's job trivial, introducing bias to the training process, so we remove single line comments from the investigated warning lines, complemented by standard pre-processing steps such as removing leading and trailing whitespace.
\paragraph{Tokenization} As pointed out by Dagan et al.~\cite{Dagan:2024} recent models use general tokenizers for specific domains such as source code analysis and completion, even though they are suboptimal for such tasks. However, high-compression tokenizers exist, such as InCoder's tokenizer~\cite{Fried:2022}, which is trained and finetuned on source code, making it more efficient than general-purpose tokenizers. Inspired by this, we use InCoder's tokenizer on the warning lines to convert them into a sequence of tokens, which are then further processed.
\paragraph{Embedding} The embedding is done by the encoder part of the transformer architecture, which is common for transformer-based embedders such as BERT~\cite{Devlin:2019}. To be more specific, we use RoFormer, a BERT-like autoencoding model with rotary position embeddings, which have shown improved performance on classification tasks with long texts~\cite{Su:2024}. We tried multiple setups of hidden size (1024, 512, 256), number of layers (1024, 512, 256) and attention heads (2, 8). We found that RoFormer with 2 layers, 512 hidden size and 8 attention heads works best for our task. To combat overfitting, we use dropout with a rate of 0.2. Both for the tokenization and embedding, we use the HuggingFace's Transformers library.\footnote{https://huggingface.co/docs/transformers/index, accessed on 2025.06.03.}
\subsection{Context Embedder (CE)}
The Context Embedder (CE, the blue rectangle in Figure~\ref{fig:architecture}) is designed to process the warning message and source code context as raw text input, parsed from the source code. All of these artifacts are processed by the mGTE sentence transformer model, introduced in Section~\ref{sec:nlp}, resulting in three separate 768-dimensional embeddings, one for each input artifact.

\paragraph{Warning message}
The warning message contains a description of the issue in natural language, so it is important to capture its semantic meaning, which is why we use mGTE to embed it. Note that using a multilingual, general purpose model comes with the advantage of being able to process warning messages in a semantic way, mitigating the risk of overfitting to a specific SCA tool's warning message format or being too specific by considering specific warning identifiers. This is in contrast to some previous works that incorporated fixed identifiers into the model~\cite{Hegedűs:2022}, restricting the approach's broader applicability.

\paragraph{Source code context}
The source code context is the lines or tokens of source code surrounding the warning line, which can provide additional information about the issue and its potential impact, as outlined in numerous works that analyze source code~\cite{Zou:2019,Li:2022,Kharkar:2022}. The number of lines is a hyperparameter of the model, which after experimenting with values of 3, 5, 25, 50, we chose 25 lines before (Pre-context) and after (Post-context) as it provided the best performance. Because the context provides general information about the warning, and also its length can vary significantly, we use mGTE to embed it as well, which is designed to handle long texts efficiently.  
\subsection{Feature Fusion (FF)}
\begin{figure*}[h]
  \centering 
  \includegraphics[width=\textwidth]{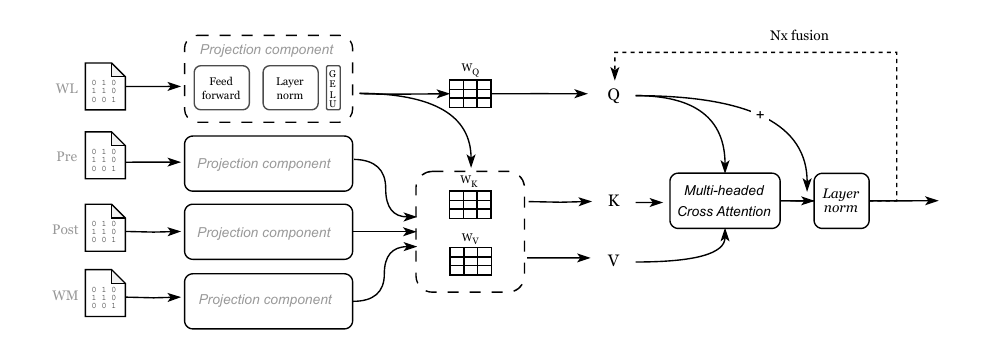}
  \caption{Feature fusion: The projected warning artifacts are mapped to Q, K, V tensors which are then used in a modular cross-attention based component}
  \label{fig:feature-fusion} 
\end{figure*}
The Feature Fusion component (FF, blue rectangle in Figure~\ref{fig:architecture}) is aimed to intelligently aggregate the context and warning line embeddings. A more in-depth breakdown of its structure can be seen in Figure~\ref{fig:feature-fusion}.

First, it uses a dedicated projection component for each feature embedding to map the different sized embeddings to a uniform size. The projection component consists of a feed forward layer, followed by layer normalization, finished with a GeLU activation function. At training time, dropout is also applied to the output of the projection layer.  

Then, the vectors are fed to a cross-attention based component to learn the relations between the different warning artifacts. To be more specific, given batch size $B$, projection dimension $P$ and embedding dimension $d_k$, the inputs are the projected vectors of \textit{Warning Line Embedding} ($\text{WL}$), \textit{Pre-Context Embedding} ($\text{Pre}$), \textit{Post-Context Embeddings} ($\text{Post}$) and \textit{Warning Message Embeddings} ($\text{WM})$ $\in \mathbb{R}^{B \times P}$. Then, the cross attention is formally defined as follows: 

\begin{itemize}
    \item $\mathbf{Q} \in \mathbb{R}^{B \times 1 \times P}$ as the \textbf{query} tensor, with $W_Q \in \mathbb{R}^{P \times d_k} $ query weight matrix.
    \item $\mathbf{K} \in \mathbb{R}^{B \times 3 \times P}$ as the \textbf{key} tensor, with $W_K \in \mathbb{R}^{P \times d_k} $ key weight matrix.
    \item $\mathbf{V} \in \mathbb{R}^{B \times 3 \times P}$ as the \textbf{value} tensor, with $W_V \in \mathbb{R}^{P \times d_k} $ value weight matrix.
\end{itemize}
Where:
\begin{itemize}
    \item The \textbf{query} tensor $\mathbf{Q}$ is derived from the projected vector $\text{WL}$ using the query weight matrix $\text{WL}_\text{q} = \text{WL} * W_Q$:
    $$ \mathbf{Q} = \left\{\text{WL}_{\text{q}}\right\} $$
    \item The \textbf{key} and \textbf{value} tensors $\mathbf{K}$ and $\mathbf{V}$ are formed by first transforming the projections $\text{WL}$, $\text{Pre}$, $\text{Post}$ and $\text{WM}$ through the key and value weight matrices, producing $\text{WL}_\text{k}, \text{Pre}_\text{k}, \text{Post}_\text{k}, \text{WM}_\text{k}$ for the keys, and similarly for the query, then concatenating them:
    $$ \mathbf{K} = \left\{\text{WL}_{\text{k}}, \text{Pre}_{\text{k}}, \text{Post}_{\text{k}}, \text{WM}_{\text{k}}\right\} $$ $$ \mathbf{V} = \left\{\text{WL}_{\text{v}}, \text{Pre}_{\text{v}}, \text{Post}_{\text{v}}, \text{WM}_{\text{v}}\right\} $$
\end{itemize}
Then the attention value is calculated as usual~\cite{Vaswani:2017}:

$$ \text{Attention}(Q, K, V) = \text{softmax}\left(\frac{QK^T}{\sqrt{d_k}}\right)V $$

Please note that the actual implementation uses attention heads, which are not formally described here for simplicity, but the overall idea is the same.

The output of the cross-attention is a tensor of shape $\mathbb{R}^{B \times 1 \times P}$, which is connected to a residual connection of the query (which initially is the projection of the warning line), followed by layer normalization. Before the final output, this cross-attention process can be repeated multiple times (as it can be observed in Figure~\ref{fig:feature-fusion}, \textit{Nx fusion} edge), to further refine the feature representation. 

\section{Research Method}

\subsection{Research Questions}
To present and interpret our results, we answer the following research questions:
\begin{itemize}
    \item \textbf{RQ1:} How effective STAF is in filtering non-actionable warnings?
    \item \textbf{RQ2:} How stable is STAF across different training runs?
    \item \textbf{RQ3:} What are the effects of the different components of the model?
\end{itemize}

\subsection{Dataset}\label{sec:dataset}
Various datasets have been used to evaluate SCA report filtering methods, however, most of them are either synthetic or limited in size and in terms of distinct warning types. For example, many early and even some recent works use the FaultBench~\cite{Heckman:2008} benchmark, which contains only 443 entries and 55 alert types. Another commonly used dataset is the OSWAP benchmark, utilized for warning filtering by Koc et al.~\cite{Koc:2019}, which contains 2371 reports, all of them being SQL injection related issues. Kharkar et al.~\cite{Kharkar:2022} used two datasets, one containing 539 null dereference and another with 108 resource leakage warnings. 

To ensure broad evaluation of our model, we used the recently published NASCAR (Non-Actionable Static Analysis Reports) dataset~\cite{Koszo:2025} to evaluate our model as it contains a large amount of SCA reports with a diverse set of warning types. It is generated by analyzing 102 GitHub repositories, which were selected based on the number of stars and activity to ensure data quality, specifically requiring at least 200 stars and updates within the last couple of years. Commits are extracted from January 1, 2022, and SCA reports are generated for each commit using PMD~\cite{PMD} and SpotBugs~\cite{SpotBugs}. The deduplicated dataset contains 1,005,778 records, with 130,011 (12.93\%) actionable and 875,767 (87.07\%) non-actionable warnings. It includes 280 different PMD bug patterns and 490 different SpotBugs bug patterns. The dataset is publicly available on Zenodo.\footnote{https://zenodo.org/records/14942251}

\subsection{Evaluation Setup and Analysis Approach}
We have evaluated our model in two different settings: within-project and cross-project. 

In the within-project setting, we split the dataset into training and test set randomly, irrespective of the projects where the alert comes from. We used an 80-20 split, where 80\% of the data was used for training and 20\% for testing. 

We have also evaluated STAF and the baselines in a cross-project setting, when the dataset is split by project, meaning that the training and test sets contain different projects, challenging the models' ability to generalize to unseen projects. We used an 80-20 split, where 80\% of the projects were used for training and 20\% for testing.

To evaluate the stability of STAF and ensure that the results are not due to random chance when splitting the dataset or initializing the model weights, we performed a stratified 5-fold cross-validation where the dataset is split into 5 folds, with each fold containing a similar distribution of actionable and non-actionable warnings. The model is trained on 4 folds and evaluated on the remaining fold, and this process is repeated 5 times, with each fold being used as the test set once in the within-project setting as it encompasses a diverse range of warning types and code contexts, providing a comprehensive basis for evaluation.

\subsection{STAF}
We implemented our model using PyTorch and HuggingFace's Transformers library. The model is trained on the NASCAR dataset, using the warning message, warning line and source code context as input. The model is trained with a batch size of 64 for 15 epochs, after which the MCC and $\text{F}_\text{1}$ score stopped improving, indicating that the model has converged. The optimizer used is Adam with a learning rate of 0.0001, and the loss function is binary cross-entropy. 
The source code and the trained model are available online on \\Zenodo: ~\url{https://zenodo.org/records/18353774}

\subsection{Baselines}
We compare our method against three baselines: DeepInferEnhance, PRISM and ChatGPT, which are described in Section~\ref{sec:related-filtering}. As the baselines differ in terms of their structure and the dataset they were originally evaluated on, we adapted them accordingly. 

\paragraph{DeepInferEnhance} 
DeepInferEnhance is a transformer-based model proposed by Kharkar et al.~\cite{Kharkar:2022} to identify false positive warnings from static analyzers. The model is a customized version of CodeBERTa, a transformer pre-trained on a large corpus of source code. While the authors of DeepInferEnhance pre-trained their own model to add C\# support, we did not require this step as our work focuses solely on Java warnings and source code, which is supported by the original CodeBERTa.

Adhering to the original methodology, we adapt the pre-trained CodeBERTa model for our classification task. As input to the model, we used the three lines of code before and after the warning line, excluding the warning line itself, as it was implemented originally. We use the tokenizer published along with CodeBERTa and add a sequence classification head to the model architecture. Our classification head, inspired by the original's 2-layer design, consists of two feed-forward layers. The input layer takes RoBERTa's hidden size, and the second layer is half the size of the first. We use a GELU activation function between the layers and a final sigmoid activation to produce a probability between 0 (non-actionable) and 1 (actionable). We finetuned the model for 20 epochs (after which the training converged) by freezing all layers except for the classification head and training it on our evaluation dataset (introduced in Section~\ref{sec:dataset}).

\paragraph{PRISM} PRISM, proposed by Yang et al.~\cite{Yang:2024}, is an approach designed to prioritize static analysis warnings by aggregating multiple Code Representation Learning (CRL) models. As the original repository for the model is no longer available, we replicated the model based solely on the descriptions provided in the paper. The core of PRISM is an ensemble method that uses a majority voting strategy to make a final prediction, with the paper exploring both hard and soft voting.
The original work tested combinations of up to nine models, of which we implemented PRISM-3, a variant utilizing the top three models. We chose this smaller combination because the performance gain for using more models was not substantial enough to justify the additional computational resources required, a trade-off noted in the original paper. Furthermore, the unavailability of the original GloVe embeddings used in the paper limited our ability to replicate the full PRISM-9 model. 
For our baseline, we implemented PRISM-3 using a hard voting mechanism. The three models we combined were FastText + BGRU, FastText + BLSTM, and Word2Vec + GRU, which are noted as the top performers in the paper. The FastText and Word2Vec embeddings were trained from scratch on our evaluation dataset, and each respective classifier head was trained for 20 epochs along with the embeddings.

\paragraph{ChatGPT} ChatGPT is based on the GPT architecture, which is a state-of-the-art large language model (LLM) designed for text generation and understanding~\cite{Brown:2020}, and it is considered as one of the most advanced LLMs. While we acknowledge that there are other high-performance commercial LLMs available, such as Google's Gemini~\cite{gemini2025} or Anthropic's Claude~\cite{claude}, the LLM landscape evolves rapidly. The relative capacity of these models for text analysis is well-represented by any of the top-tier options, so we selected ChatGPT as our baseline because it is arguably the most prominent and widely adopted LLM. Even though ChatGPT is primarily designed for text generation, it has been shown to also perform well on other NLP tasks, including classification~\cite{Wen:2024, Li:2024}.

To use ChatGPT for our classification task, we used the prompt shown in Figure~\ref{fig:llm_prompt}. It instructs the model to classify the warning as actionable or non-actionable based on the warning message, warning line and source code context. As per best practices, we used few shot prompting, providing the model with a few examples of actionable and non-actionable warnings to guide its classification.

We classified 100 randomly sampled warnings from the dataset with a distribution of actionable and non-actionable warnings that follows the complete dataset's distribution, meaning 80 actionable and 20 non-actionable warnings. As part of the few-shot setting, we tried multiple setups and found that the model performed best with 10 examples with 6 lines of context, and 15 examples with 6 lines of context for the within-project and cross-project settings, respectively (for details, please refer to~\ref{sec:results}). We used the GPT-5 model through the OpenAI API.

\subsection{Performance Indicators}

    \paragraph{Accuracy} Represents the proportion of total correct predictions among the total number of cases examined.
    \begin{equation}
        Accuracy = \frac{TP + TN}{TP + TN + FP + FN}
    \end{equation}

    \paragraph{Precision} Also known as Positive Predictive Value, it measures the accuracy of positive predictions.
    \begin{equation}
        Precision = \frac{TP}{TP + FP}
    \end{equation}

    \paragraph{Recall} Also referred to as Sensitivity or True Positive Rate, it measures the ability of the model to find all relevant cases within a dataset.
    \begin{equation}
        Recall = \frac{TP}{TP + FN}
    \end{equation}

    \paragraph{F$_1$ Score} The harmonic mean of Precision and Recall, providing a single metric to balance the trade-off between the two.
    \begin{equation}
        F_1 = 2 \cdot \frac{Precision \cdot Recall}{Precision + Recall}
    \end{equation}

    \paragraph{Matthews Correlation Coefficient (MCC)} A correlation coefficient between the observed and predicted binary classifications that yields a value between -1 and +1. MCC is calculated as 
    \begin{equation}
       \frac{TP \cdot TN - FP \cdot FN}{\sqrt{(TP + FP)(TP + FN)(TN + FP)(TN + FN)}}
    \end{equation}

While both $F_1$ and $MCC$ are used to assess binary classification, they differ in how they handle dataset imbalance.

Briefly, the $F_1$ score is asymmetric, it focuses primarily on the positive class ($TP$) and ignores the performance on the negative class ($TN$). Consequently, a high $F_1$ score does not guarantee that the model performs well on negative instances. In contrast, $MCC$ is a symmetric measure that considers all four quadrants of the confusion matrix equally, therefore, it can be considered more informative when dealing with imbalanced datasets.

\begin{figure}[htbp]
    \centering
    \begin{lstlisting}[
        basicstyle=\ttfamily\small, % Use a small typewriter font
        breaklines=true,          % Automatically wrap long lines
        frame=single,             % Draw a single-line box around the code
        language={},              % Tell listings this isn't a programming language
        breakindent=0pt,              
        xleftmargin=\parindent    % Indent the box slightly
    ]
You are a senior developer performing an as in-depth code review as he possibly can.

You are given a code snippet and a warning message a static analyzer tool provided as explanation as to why the code snippet has possible coding problems.

Your job is to determine whether the warning is about an actionable or a non-actionable problem.

To help you decide, here are some examples (True means it is actionable, False means it is non-actionable):

Warning message: [example message]
Code: [example code]
Answer: [example answer]

Warning message: [example message]
Code: [example code]
Answer: [example answer]

...

Now here is the warning message and the code you have to examine:
Warning message: [message]
Code: [code]

Do you think the problem is actionable or non-actionable?
Your answer should only contain TRUE, if the problem is actionable, and only contain FALSE, if the problem is non-actionable.
\end{lstlisting}
    \caption{The full prompt given to the LLM to classify static analysis warnings}
    \label{fig:llm_prompt}
\end{figure}

\section{Results and Discussion}
\label{sec:results}

\subsection{How effective STAF is in filtering non-actionable warnings?}

We have evaluated our model in two different settings: within-project and cross-project. The former setting measures the model's ability to learn from and generalize to data across all projects, while the latter setting challenges the model to generalize to completely unseen projects.

\paragraph{Within-project}
\begin{table}[htb!]
	\centering
    \caption{Performance metrics for the baselines and our proposed model in a "within-project" setting}
	\begin{tabular}{lcccccc} 
		\toprule
		Model & Accuracy & MCC & $\text{F}_\text{1}$& Precision & Recall \\ 
		\midrule
		\bf{STAF} & \bf{0.97} & \bf{0.87} & \bf{0.89} & 0.91 & \bf{0.87} \\ 
		PRISM-3 & 0.95 & 0.83 & 0.78 & \bf{0.94} & 0.66 \\ 
		DeepInferEnhance & 0.93  & 0.63 & 0.64 & 0.85 & 0.51 \\ 
		ChatGPT & 0.5 & 0.08 & 0.27 & 0.17 & 0.64 \\ 
		\bottomrule
	\end{tabular}
	\vspace{1mm}
	\label{tab:results}
\end{table}
The results of the within-project evaluation can be seen in Table~\ref{tab:results}. Regarding accuracy, all models perform similarly well, with STAF achieving 0.97, which is slightly higher than the baselines. However, accuracy alone does not provide a complete picture of the models' performance, as the dataset is highly imbalanced, mirroring the real-world scenario where non-actionable warnings are much more common than actionable ones. Therefore, other metrics such as MCC, $\text{F}_\text{1}$ score, precision and recall can be more informative. 

STAF achieves a MCC of 0.87 which is higher than all of the baselines, indicating that it is effective in filtering non-actionable warnings when considering all of TP, TN, FP and FN values. The $\text{F}_\text{1}$ score of 0.89 also indicates that STAF is effective in balancing precision and recall, which is important in the context of SCA reports, as we want to minimize the number of false positives while still capturing as many true positives as possible. 

In terms of precision, STAF is slightly outperformed by PRISM-3 with a 3\% difference, at a cost of lower recall (by 22\%), meaning that PRISM-3 is more likely to miss actionable warnings.

\paragraph{Cross-project}
\begin{table}[htb!]
	\centering
	\caption{Performance metrics for the baselines and our proposed model in a "cross-project" setting}
	\begin{tabular}{lcccccc} 
		\toprule
		Model & Accuracy & MCC & $\text{F}_\text{1}$& Precision & Recall \\ 
		\midrule
		\bf{STAF} & 0.80 & \bf{0.21} & \bf{0.31} & \bf{0.41} & 0.25 \\ 
		PRISM-3 & \bf{0.81} & 0.12 & 0.15 & 0.39 & 0.09 \\ 
		DeepInferEnhance &  0.78 & 0.01 & 0.11 & 0.19 & 0.08 \\ 
		ChatGPT & 0.52 & 0.06 & 0.25 & 0.16 & \bf{0.57} \\ 
		\bottomrule
	\end{tabular}
	\vspace{1mm}
	\label{tab:results-project-split}
\end{table}
Table~\ref{tab:results-project-split} shows the results of the cross-project evaluation. As expected, the performance of all models drops compared to the within-project setting as generalizing to unseen projects is a more challenging task.
In terms of MCC, STAF significantly outperforms all other models with a score of 0.21 with an edge of 9\% over PRISM-3, while DeepInferEnhance and ChatGPT perform extremely poorly. All of the models have a better performance in terms of $\text{F}_\text{1}$ score but STAF still outperforms the baselines. Since the $\text{F}_\text{1}$ score primarily measures performance on the positive class while MCC provides a balanced score across all four confusion-matrix categories, this suggests that the baseline models struggle with correctly identifying negative instances (True Negatives).

Note that While PRISM-3 has a slightly better accuracy (by 1\%), it is outperformed by STAF in all other, imbalance sensitive metrics. More notable is the performance of ChatGPT, which has a recall score of 0.57 - a considerably higher score than all of the other models, at the cost of the lowest precision. 

\begin{answerbox}
    \noindent \textbf{Answer to RQ1}
    
    STAF demonstrates high effectiveness compared to baseline models in distinguishing actionable from non-actionable warnings. This effectiveness is observed in both within-project and cross-project experimental setups, indicating that the approach captures robust patterns of actionability while remaining accurate in project-specific contexts.
\end{answerbox}

\subsection{How stable is STAF across different training runs?}
\begin{table}[htb!]
    \centering
    \caption{Cross-validation results on 5 folds}
    \begin{tabular}{lccccc}
        \toprule
        Model & Accuracy & MCC & $\text{F}_\text{1}$& Precision & Recall \\
        \midrule
        Fold 1 & 0.97 & 0.87 & 0.88 & 0.92 & 0.85 \\
        Fold 2 & 0.97 & 0.87 & 0.89 & 0.93 & 0.85 \\
        Fold 3 & 0.97 & 0.87 & 0.89 & 0.90 & 0.87 \\
        Fold 4 & 0.97 & 0.87 & 0.88 & 0.95 & 0.82 \\
        Fold 5 & 0.97 & 0.87 & 0.89 & 0.91 & 0.87 \\
        \bottomrule
    \end{tabular}
    \vspace{1mm}
    \label{tab:xval}
\end{table}
The results of the cross-validation are presented in Table~\ref{tab:xval}. It is evident that the results are consistent across the folds as accuracy and MCC is unchanged, while $\text{F}_\text{1}$ score, precision and recall shows a negligible deviation.

\begin{answerbox}
    \noindent \textbf{Answer to RQ2}
    
    The stratified 5-fold cross-validation in the within-project setting confirms that the model's performance is consistent with minimal variance attributable to data splitting or initialization.
\end{answerbox}

\subsection{What are the effects of the different components of the model?} 
\begin{table}[htb!]
	\centering
	\caption{Performance metrics for STAF and the ablation study models}
	\begin{tabular}{lcccccc} 
		\toprule
		Model & Accuracy & MCC & $\text{F}_\text{1}$& Precision & Recall \\ 
		\midrule
		\bf{STAF} & \bf{0.97} & \bf{0.87} & \bf{0.89} & \bf{0.91} & \bf{0.88} \\ 
		STAF-PRECTX & 0.96 & 0.84 & 0.85 & 0.90 & 0.81 \\ 
		STAF-POSTCTX & \bf{0.97} & 0.85 & 0.87 & 0.89 & 0.85 \\ 
		STAF-CTX & 0.95 & 0.76 & 0.79 & 0.81 & 0.77 \\ 
		STAF-WMSG & \bf{0.97} & 0.84 & 0.86 & 0.89 & 0.83 \\ 
		\bottomrule
	\end{tabular}
	\vspace{1mm}
	\label{tab:ablation}
\end{table}
To answer this question, we performed an ablation study where we systematically removed components from the model and evaluated their impact on the performance metrics. In Table~\ref{tab:ablation} we present the results, where we evaluate the effect of the different components of our model in a within-project setting. We compare the full model (STAF) against four variants: STAF-PRECTX, STAF-POSTCTX, STAF-CTX and STAF-WMSG, in which we remove the pre-context, post-context, both pre and post context and warning message embeddings, respectively. All of the models are trained on the same dataset for 15 epochs as the full model and evaluated using the same metrics as in Table~\ref{tab:results}.
The results can be seen in Table~\ref{tab:ablation}. It can be seen that the full model (STAF) outperforms all variants in terms of all metrics, indicating that all components are important for the model to achieve the best result. However, the impact of the different components varies. 

Removing the pre-context or post-context embeddings has a slightly negative impact on the model's performance, with a 3\%, 2\% drop in MCC and 4\%, 2\% drop in $\text{F}_\text{1}$ score, respectively. With both contexts removed the performance is affected significantly: MCC drops by 11\% and $\text{F}_\text{1}$ score by 10\%. This indicates that considering either the pre-context or the post-context is important, while using them together provides the best result. 

Removing the warning message embedding has a smaller impact on the performance, with a 3\% drop in MCC and $\text{F}_\text{1}$ score, showing that although the warning message has less impact on the overall performance than the context embeddings, it is still important.
\begin{answerbox}
    \noindent \textbf{Answer to RQ3}
    
    The inclusion of code context is critical for classification performance. Our results indicate that models utilizing full context features outperform the other setups, confirming that the environment is essential for identifying actionability.
\end{answerbox}

\section{Threats to Validity}

In this section, we discuss the potential threats to the validity of our study and the steps we have taken to mitigate them.

\paragraph{Overfitting}
A primary threat to internal validity concerns hyperparameter selection. While we experimented with various configurations for context length and model architecture, there remains a risk that our choices are optimized specifically for the NASCAR dataset~\cite{Koszo:2025}. However, most related works in this field utilize significantly smaller benchmarks, such as FaultBench (443 entries) \cite{Heckman:2008} or the dataset used by Kharkar et al. (539 warnings) \cite{Kharkar:2022}. Because STAF is trained and evaluated on more than 1 million records, the results are inherently more general and less prone to overtuning typical of smaller-scale studies. Nevertheless, a more exhaustive search of the hyperparameter space could further refine these results.

\paragraph{Language Specificity}
Regarding the generalizability of our findings, language specificity is a notable constraint. Our current evaluation is restricted to Java projects. We purposefully restricted our focus to the Java ecosystem because Java and C/C++ are the most frequently supported programming languages among existing Static Code Analysis (SCA) tools \cite{Guo:2023}. Furthermore, Java is the most extensively studied language in the related literature on false positive (FP) mitigation, with the academic community predominantly proposing novel mitigation approaches for Java-based SA tools \cite{Guo:2023}. While Java provides a massive ecosystem and a strong baseline for static analysis research, we acknowledge that the patterns of actionability we identified may differ in other practically popular languages like Python or C++. Expanding STAF to a language-agnostic framework is an objective of our future work.

\paragraph{Actionability Definition}
A common challenge in SCA research is defining actionability. Actionability can be subjective and vary between individual developers. To mitigate this, we rely on the labeling logic of the NASCAR dataset, where a warning is classified as actionable only if it is addressed through subsequent commits in well-maintained, high quality projects. This approach suggests that the findings reflect project wide maintenance decisions rather than personal preferences. 

\paragraph{Project Leakage}
Finally, we acknowledge the risk of project leakage. In large-scale mining of GitHub repositories, it is possible that code snippets or library dependencies are shared across different projects, which could subtly influence the cross-project evaluation. Although this is a difficult challenge to address in practice, we plan to investigate techniques to ensure stricter isolation during training.

\section{Conclusion and Future Work}

In this paper, we addressed the issue of alert fatigue in static code analysis by proposing STAF: Sentence Transformer-based Actionability Filtering. Using a transformer-based architecture that combines a dedicated Warning Line Embedder (WLE) with a general-purpose Context Embedder (CE), our model effectively captures the semantic nuances of warning messages, raw source code, and their surrounding context. 

Our evaluation on the NASCAR dataset, comprising over 1 million reports from Java projects, demonstrates that STAF significantly outperforms state-of-the-art baselines. Specifically, in a within-project setting, our approach reached an $F_{1}$ score of 0.89 and an MCC of 0.87, outperforming existing methods for SCA warning filtering by at least 11\%. Furthermore, STAF exhibited superior generalization capabilities in cross-project evaluations, maintaining higher reliability in the face of the class imbalance inherent in real-world static analysis reports.

Through the ablation study we confirmed that the fusion of pre-context, post-context, and warning line information is critical for achieving optimal performance. The results suggest that while the warning line remains the central artifact, the surrounding source code provides vital evidence for determining actionability. By reducing the volume of non-actionable findings, STAF offers a scalable solution to enhance the usability of SCA tools and developer productivity.

In terms of future work, we aim to integrate STAF into widely adopted Quality Assurance (QA) platforms, such as SonarQube, to evaluate its practical utility and impact on alert fatigue within industrial developer workflows. Furthermore, we plan to expand our research beyond Java by applying STAF to other programming languages. This includes investigating the potential of a language-agnostic setup to determine if the context-aware patterns of warning actionability can be effectively generalized. We would also like to investigate more thoroughly the possibility of incorporating project-wide metadata into each warning, so that the actionability filtering pipeline can be tailored for specific projects more effectively.

\section{Limitations}
One of the primary limitations of our current approach is that it treats alert filtering as a strictly binary classification problem (Actionable vs. Non-Actionable) and does not incorporate standard severity labels (e.g., "Critical," "Major," "Minor," or "Info") typically provided by static code analysis tools~\cite{sonarqube, SpotBugs, PMD}. Also, note that a developer's decision to act on a warning in industrial practice is often a risk assessment rather than a purely syntactic decision. Consequently, there is a theoretical safety risk that the model might correctly learn to classify a "Minor" code smell as non-actionable (because developers frequently ignore it to save time), but then generate a dangerous False Negative by applying that same logic to a "Critical" security vulnerability that happens to appear syntactically similar to the ignored pattern.

Despite this theoretical risk, we intentionally excluded tool-provided severity labels to ensure that STAF remains a highly generalizable solution. Different SCA tools employ vastly different categories, rules, and taxonomies for grading severity. Accounting for these specific, tool-dependent labels would tightly couple our model to the distinct outputs of PMD or SpotBugs, making the solution rigid and difficult to expand to other tools or programming languages in the future. 

Furthermore, the risk of conflating minor issues with critical vulnerabilities is typically minimal in practice. Minor code smells (such as stylistic issues or suboptimal variable declarations) and critical security vulnerabilities (such as SQL injections or resource leaks) rely on fundamentally different logic and code structures. Because our model deeply analyzes the source code context and warning lines, these issues are rarely syntactically similar enough for the embeddings to confuse a critical vulnerability with a benign code smell.

Our dataset curation strategy also serves as a strong mitigating factor against this threat. The model is trained on the NASCAR dataset, which strictly consists of well-maintained, high-quality projects (requiring at least 200 stars and recent update activity, please refer to Section \ref{sec:dataset}). In the dataset, a warning is labeled as non-actionable (a False Positive) only if it persists unaddressed for the entire observed time window~\cite{Koszo:2025}. The probability of an actually critical security vulnerability remaining ignored for an extended period in such highly scrutinized, actively maintained projects is low. 

Finally, while the combination of contextual embedding differences and strict dataset quality control makes the chance of this issue occurring minimal, it must be acknowledged that this limitation can still possibly manifest. Because the safety implications of a false negative for a critical security flaw are severe, our binary classification approach is best utilized as a powerful prioritization aid to reduce alert fatigue, rather than a total replacement for human-driven security risk assessments.

\section*{Acknowledgment}
This work was supported by the Ministry of Culture and Innovation of Hungary from the National Research, Development and Innovation Fund, through the K\_23, OTKA Funding Scheme, under Project K 147225.
The work has also received support from the European Union Horizon Program under the grant number 101120393 (Sec4AI4Sec). 

\bibliographystyle{ACM-Reference-Format}
% \clearpage
\balance
\bibliography{refs}

@misc{gemini2025,
      title={Gemini: A Family of Highly Capable Multimodal Models}, 
      author={Gemini Team and Rohan Anil and Sebastian Borgeaud et al.},
      year={2025},
      eprint={2312.11805},
      archivePrefix={arXiv},
      primaryClass={cs.CL},
      url={https://arxiv.org/abs/2312.11805}, 
}

@inproceedings{Heckman:2009,
author = {Heckman, Sarah and Williams, Laurie},
year = {2009},
month = {04},
pages = {161-170},
title = {A Model Building Process for Identifying Actionable Static Analysis Alerts},
doi = {10.1109/ICST.2009.45}
}

@article{Guo:2023,
author = {Guo, Zhaoqiang and Tan, Tingting and Liu, Shiran and Liu, Xutong and Lai, Wei and Yang, Yibiao and Li, Yanhui and Chen, Lin and Dong, Wei and Zhou, Yuming},
title = {Mitigating False Positive Static Analysis Warnings: Progress, Challenges, and Opportunities},
year = {2023},
issue_date = {Dec. 2023},
publisher = {IEEE Press},
volume = {49},
number = {12},
issn = {0098-5589},
url = {https://doi.org/10.1109/TSE.2023.3329667},
doi = {10.1109/TSE.2023.3329667},
journal = {IEEE Trans. Softw. Eng.},
month = dec,
pages = {5154–5188},
numpages = {35}
}

@inproceedings{Balachandran:2013,
   AUTHOR     = {Balachandran, Vipin},
   TITLE      = {Reducing human effort and improving quality in peer code reviews using automatic static analysis and reviewer recommendation},
   YEAR       = {2013},
   ISBN       = {9781467330763},
   PUBLISHER  = {IEEE Press},
   PAGES      = {931–940},
   NUMPAGES   = {10},
   LOCATION   = {San Francisco, CA, USA},
   SERIES     = {ICSE '13}
}

@article{Bessey:2010,
   AUTHOR     = {Bessey, Al
               AND Block, Ken
               AND Chelf, Ben
               AND Chou, Andy
               AND Fulton, Bryan
               AND Hallem, Seth
               AND Henri-gros, Charles
               AND Kamsky, Asya
               AND Mcpeak, Scott
               AND Engler, Dawson},
   TITLE      = {A few billion lines of code later: using static analysis to find bugs in the real world},
   YEAR       = {2010},
   ISSUE_DATE = {February 2010},
   PUBLISHER  = {Association for Computing Machinery},
   ADDRESS    = {New York, NY, USA},
   VOLUME     = {53},
   NUMBER     = {2},
   ISSN       = {0001-0782},
   URL        = {https://doi.org/10.1145/1646353.1646374},
   DOI        = {10.1145/1646353.1646374},
   JOURNAL    = {Commun. ACM},
   MONTH      = feb,
   PAGES      = {66–75},
   NUMPAGES   = {10}
}

@article{Heckman:2011,
   TITLE      = {A systematic literature review of actionable alert identification techniques for automated static code analysis},
   JOURNAL    = {Information and Software Technology},
   VOLUME     = {53},
   NUMBER     = {4},
   PAGES      = {363-387},
   YEAR       = {2011},
   NOTE       = {Special section: Software Engineering track of the 24th Annual Symposium on Applied Computing},
   ISSN       = {0950-5849},
   DOI        = {https://doi.org/10.1016/j.infsof.2010.12.007},
   URL        = {https://www.sciencedirect.com/science/article/pii/S0950584910002235},
   AUTHOR     = {Heckman, Sarah AND Williams Laurie},
}

@inproceedings{Kharkar:2022,
   AUTHOR     = {Kharkar, Anant
               AND Moghaddam, Roshanak Zilouchian
               AND Jin, Matthew
               AND Liu, Xiaoyu
               AND Shi, Xin
               AND Clement, Colin
               AND Sundaresan, Neel},
   TITLE      = {Learning to reduce false positives in analytic bug detectors},
   YEAR       = {2022},
   ISBN       = {9781450392211},
   PUBLISHER  = {Association for Computing Machinery},
   ADDRESS    = {New York, NY, USA},
   URL        = {https://doi.org/10.1145/3510003.3510153},
   DOI        = {10.1145/3510003.3510153},
   BOOKTITLE  = {Proceedings of the 44th International Conference on Software Engineering},
   PAGES      = {1307–1316},
   NUMPAGES   = {10},
   LOCATION   = {Pittsburgh, Pennsylvania},
   SERIES     = {ICSE '22}
}

@inproceedings{Ayewah:2007,
   AUTHOR     = {Ayewah, Nathaniel
               AND Pugh, William
               AND Morgenthaler, J.
               AND Penix, John
               AND Zhou, Yuqian},
   YEAR       = {2007},
   MONTH      = {06},
   PAGES      = {1-8},
   TITLE      = {Evaluating static analysis defect warnings on production software},
   DOI        = {10.1145/1251535.1251536}
}

@INPROCEEDINGS{Baca:2010,
   AUTHOR     = {Baca, Dejan},
   BOOKTITLE  = {2010 International Conference on Availability, Reliability and Security}, 
   TITLE      = {Identifying Security Relevant Warnings from Static Code Analysis Tools through Code Tainting}, 
   YEAR       = {2010},
   VOLUME     = {},
   NUMBER     = {},
   PAGES      = {386-390},
   DOI        = {10.1109/ARES.2010.108}}

@INPROCEEDINGS{Lee:2019,
   AUTHOR     = {Lee, Seongmin
               AND Hong, Shin
               AND Yi, Jungbae
               AND Kim, Taeksu
               AND Kim, Chul-joo
               AND Yoo, Shin},
   BOOKTITLE  = {2019 12th IEEE Conference on Software Testing, Validation and Verification (ICST)}, 
   TITLE      = {Classifying False Positive Static Checker Alarms in Continuous Integration Using Convolutional Neural Networks}, 
   YEAR       = {2019},
   VOLUME     = {},
   NUMBER     = {},
   PAGES      = {391-401},
   DOI        = {10.1109/ICST.2019.00048}}

@ARTICLE{Hegedűs:2022,
   AUTHOR     = {Hegedűs, Péter AND Ferenc, Rudolf},
   JOURNAL    = {IEEE Access}, 
   TITLE      = {Static Code Analysis Alarms Filtering Reloaded: A New Real-World Dataset and its ML-Based Utilization}, 
   YEAR       = {2022},
   VOLUME     = {10},
   NUMBER     = {},
   PAGES      = {55090-55101},
   DOI        = {10.1109/ACCESS.2022.3176865}}

@article{Su:2024,
   AUTHOR     = {Su, Jianlin
               AND Ahmed, Murtadha
               AND Lu, Yu
               AND Pan, Shengfeng
               AND Bo, Wen
               AND Liu, Yunfeng},
   TITLE      = {RoFormer: Enhanced transformer with Rotary Position Embedding},
   YEAR       = {2024},
   ISSUE_DATE = {Feb 2024},
   PUBLISHER  = {Elsevier Science Publishers B. V.},
   ADDRESS    = {NLD},
   VOLUME     = {568},
   NUMBER     = {C},
   ISSN       = {0925-2312},
   URL        = {https://doi.org/10.1016/j.neucom.2023.127063},
   DOI        = {10.1016/j.neucom.2023.127063},
   JOURNAL    = {Neurocomput.},
   MONTH      = feb,
   NUMPAGES   = {12}
}

@inproceedings{reimers-2020-multilingual-sentence-bert,
  title = "Making Monolingual Sentence Embeddings Multilingual using Knowledge Distillation",
  author = "Reimers, Nils and Gurevych, Iryna",
  booktitle = "Proceedings of the 2020 Conference on Empirical Methods in Natural Language Processing",
  month = "11",
  year = "2020",
  publisher = "Association for Computational Linguistics",
  url = "https://arxiv.org/abs/2004.09813",
}

@inproceedings{Zhang:2024,
   TITLE      = {mGTE: Generalized Long-Context Text Representation and Reranking Models for Multilingual Text Retrieval},
   AUTHOR     = {Zhang, Xin
               AND Zhang, Yanzhao
               AND Long, Dingkun
               AND Xie, Wen
               AND Dai, Ziqi
               AND Tang, Jialong
               AND Lin, Huan
               AND Yang, Baosong
               AND Xie, Pengjun
               AND Huang, Fei
               AND Others},
   BOOKTITLE  = {Proceedings of the 2024 Conference on Empirical Methods in Natural Language Processing: Industry Track},
   PAGES      = {1393--1412},
   YEAR       = {2024}
}

@online{claude,
	title = {Anthropic Claude - Accessed 2026-01-15},
	url = {https://claude.ai},
}

@article{Fried:2022,
   TITLE      = {Incoder: A generative model for code infilling and synthesis},
   AUTHOR     = {Fried, Daniel
               AND Aghajanyan, Armen
               AND Lin, Jessy
               AND Wang, Sida
               AND Wallace, Eric
               AND Shi, Freda
               AND Zhong, Ruiqi
               AND Yih, Wen-tau
               AND Zettlemoyer, Luke
               AND Lewis, Mike},
   JOURNAL    = {arXiv preprint arXiv:2204.05999},
   YEAR       = {2022}
}

@inproceedings{Dagan:2024,
   AUTHOR     = {Dagan, Gautier
               AND Synnaeve, Gabriel
               AND Rozi\`{e}re, Baptiste},
   TITLE      = {Getting the most out of your tokenizer for pre-training and domain adaptation},
   YEAR       = {2024},
   PUBLISHER  = {JMLR.org},
   BOOKTITLE  = {Proceedings of the 41st International Conference on Machine Learning},
   ARTICLENO  = {387},
   NUMPAGES   = {22},
   LOCATION   = {Vienna, Austria},
   SERIES     = {ICML'24}
}

@inproceedings{Devlin:2019,
   AUTHOR     = {Devlin, Jacob
               AND Chang, Ming{-}wei 
               AND Lee, Kenton 
               AND Toutanova, Kristina },
   EDITOR     = {Jill Burstein and
                  Christy Doran and
                  Thamar Solorio},
   TITLE      = {{BERT:} Pre-training of Deep Bidirectional Transformers for Language
                  Understanding},
   BOOKTITLE  = {Proceedings of the 2019 Conference of the North American Chapter of
                  the Association for Computational Linguistics: Human Language Technologies,
                  {NAACL-HLT} 2019, Minneapolis, MN, USA, June 2-7, 2019, Volume 1 (Long
                  and Short Papers)},
   PAGES      = {4171--4186},
   PUBLISHER  = {Association for Computational Linguistics},
   YEAR       = {2019},
   URL        = {https://doi.org/10.18653/v1/n19-1423},
   DOI        = {10.18653/V1/N19-1423},
   BIBSOURCE  = {dblp computer science bibliography, https://dblp.org}
}

@inproceedings{Wei:2017,
   AUTHOR     = {Wei, Lili
               AND Liu, Yepang
               AND Cheung, Shing-chi},
   TITLE      = {OASIS: prioritizing static analysis warnings for Android apps based on app user reviews},
   YEAR       = {2017},
   ISBN       = {9781450351058},
   PUBLISHER  = {Association for Computing Machinery},
   ADDRESS    = {New York, NY, USA},
   URL        = {https://doi.org/10.1145/3106237.3106294},
   DOI        = {10.1145/3106237.3106294},
   BOOKTITLE  = {Proceedings of the 2017 11th Joint Meeting on Foundations of Software Engineering},
   PAGES      = {672–682},
   NUMPAGES   = {11},
   LOCATION   = {Paderborn, Germany},
   SERIES     = {ESEC/FSE 2017}
}

@inproceedings{Kim:2007,
   AUTHOR     = {Kim, Sunghun AND Ernst, Michael D.},
   TITLE      = {Which warnings should I fix first?},
   YEAR       = {2007},
   ISBN       = {9781595938114},
   PUBLISHER  = {Association for Computing Machinery},
   ADDRESS    = {New York, NY, USA},
   URL        = {https://doi.org/10.1145/1287624.1287633},
   DOI        = {10.1145/1287624.1287633},
   BOOKTITLE  = {Proceedings of the the 6th Joint Meeting of the European Software Engineering Conference and the ACM SIGSOFT Symposium on The Foundations of Software Engineering},
   PAGES      = {45–54},
   NUMPAGES   = {10},
   LOCATION   = {Dubrovnik, Croatia},
   SERIES     = {ESEC-FSE '07}
}

@InProceedings{Chen:2013,
author="Chen, Chen
and Lu, Kai
and Wang, Xiaoping
and Zhou, Xu
and Fang, Li",
editor="Wu, Chenggang
and Cohen, Albert",
title="Pruning False Positives of Static Data-Race Detection via Thread Specialization",
booktitle="Advanced Parallel Processing Technologies",
year="2013",
publisher="Springer Berlin Heidelberg",
address="Berlin, Heidelberg",
pages="77--90",
isbn="978-3-642-45293-2"
}

@INPROCEEDINGS{Muske:2015,
   AUTHOR     = {Muske, Tukaram AND Khedker, Uday P.},
   BOOKTITLE  = {2015 IEEE 26th International Symposium on Software Reliability Engineering (ISSRE)}, 
   TITLE      = {Efficient elimination of false positives using static analysis}, 
   YEAR       = {2015},
   VOLUME     = {},
   NUMBER     = {},
   PAGES      = {270-280},
   DOI        = {10.1109/ISSRE.2015.7381820}}

@InProceedings{Valdiviezo:2014,
author="Valdiviezo, Manuel
and Cifuentes, Cristina
and Krishnan, Padmanabhan",
editor="Garrigue, Jacques",
title="A Method for Scalable and Precise Bug Finding Using Program Analysis and Model Checking",
booktitle="Programming Languages and Systems",
year="2014",
publisher="Springer International Publishing",
address="Cham",
pages="196--215",
isbn="978-3-319-12736-1"
}

@INPROCEEDINGS{Koc:2019,
   AUTHOR     = {Koc, Ugur
               AND Wei, Shiyi
               AND Foster, Jeffrey S.
               AND Carpuat, Marine
               AND Porter, Adam A.},
   BOOKTITLE  = {2019 12th IEEE Conference on Software Testing, Validation and Verification (ICST)}, 
   TITLE      = {An Empirical Assessment of Machine Learning Approaches for Triaging Reports of a Java Static Analysis Tool}, 
   YEAR       = {2019},
   VOLUME     = {},
   NUMBER     = {},
   PAGES      = {288-299},
   DOI        = {10.1109/ICST.2019.00036}}

@inproceedings{Yang:2024,
  	ADDRESS    = {Rovaniemi, Finland},
  	TITLE      = {Reducing {False} {Positives} of {Static} {Bug} {Detectors} {Through} {Code} {Representation} {Learning}},
  	COPYRIGHT  = {https://doi.org/10.15223/policy-029},
  	ISBN       = {979-8-3503-3066-3},
  	URL        = {https://ieeexplore.ieee.org/document/10589733/},
  	DOI        = {10.1109/SANER60148.2024.00075},
  	LANGUAGE   = {en},
  	URLDATE    = {2025-02-28},
  	BOOKTITLE  = {2024 {IEEE} {International} {Conference} on {Software} {Analysis}, {Evolution} and {Reengineering} ({SANER})},
  	PUBLISHER  = {IEEE},
  	AUTHOR     = {Yang, Yixin
               AND Wen, Ming
               AND Gao, Xiang
               AND Zhang, Yuting
               AND Sun, Hailong},
  	MONTH      = mar,
  	YEAR       = {2024},
  	PAGES      = {681--692},
}

@misc{Reimers:2019,
  	TITLE      = {Sentence-{BERT}: {Sentence} {Embeddings} using {Siamese} {BERT}-{Networks}},
  	SHORTTITLE = {Sentence-{BERT}},
  	URL        = {http://arxiv.org/abs/1908.10084},
  	DOI        = {10.48550/arXiv.1908.10084},
  	LANGUAGE   = {en},
  	URLDATE    = {2025-06-12},
  	PUBLISHER  = {arXiv},
  	AUTHOR     = {Reimers, Nils AND Gurevych, Iryna},
  	MONTH      = aug,
  	YEAR       = {2019},
  	NOTE       = {arXiv:1908.10084 [cs]},
}

@article{Mikolov:2013,
   AUTHOR     = {Mikolov, Tomas
               AND Chen, Kai
               AND Corrado, G.s
               AND Dean, Jeffrey},
   YEAR       = {2013},
   MONTH      = {01},
   PAGES      = {},
   TITLE      = {Efficient Estimation of Word Representations in Vector Space},
   VOLUME     = {2013},
   JOURNAL    = {Proceedings of Workshop at ICLR}
}

@InProceedings{Joulin:2017,
   TITLE      = {Bag of Tricks for Efficient Text Classification},
   AUTHOR     = {Joulin, Armand
               AND Grave, Edouard
               AND Bojanowski, Piotr
               AND Mikolov, Tomas},
   BOOKTITLE  = {Proceedings of the 15th Conference of the European Chapter of the Association for Computational Linguistics: Volume 2, Short Papers},
   MONTH      = {April},
   YEAR       = {2017},
   PUBLISHER  = {Association for Computational Linguistics},
   PAGES      = {427--431},
}

@unknown{Liu:2019,
   AUTHOR     = {Liu, Yinhan
               AND Ott, Myle
               AND Goyal, Naman
               AND Du, Jingfei
               AND Joshi, Mandar
               AND Chen, Danqi
               AND Levy, Omer
               AND Lewis, Mike
               AND Zettlemoyer, Luke
               AND Stoyanov, Veselin},
   YEAR       = {2019},
   MONTH      = {07},
   PAGES      = {},
   TITLE      = {RoBERTa: A Robustly Optimized BERT Pretraining Approach},
   DOI        = {10.48550/arXiv.1907.11692},
}

@inproceedings{Feng:2020,
   AUTHOR     = {Zhangyin, Feng
               AND Daya, Guo
               AND Duyu, Tang
               AND Nan, Duan
               AND Xiaocheng, Feng
               AND Ming, Gong
               AND Linjun, Shou
               AND Bing, Qin
               AND Ting, Liu
               AND Daxi,n Jiang
               AND Ming, Zhou},
   EDITOR     = {Trevor Cohn and
                  Yulan He and
                  Yang Liu},
   TITLE      = {CodeBERT: {A} Pre-Trained Model for Programming and Natural Languages},
   BOOKTITLE  = {Findings of the Association for Computational Linguistics: {EMNLP}
                  2020, Online Event, 16-20 November 2020},
   SERIES     = {Findings of {ACL}},
   VOLUME     = {{EMNLP} 2020},
   PAGES      = {1536--1547},
   PUBLISHER  = {Association for Computational Linguistics},
   YEAR       = {2020},
   URL        = {https://doi.org/10.18653/v1/2020.findings-emnlp.139},
   DOI        = {10.18653/V1/2020.FINDINGS-EMNLP.139},
   BIBSOURCE  = {dblp computer science bibliography, https://dblp.org}
}

@article{Muennighoff:2022,
   AUTHOR     = {Muennighoff, Niklas
               AND Tazi, Nouamane
               AND Magne, Lo{\"\i}c
               AND Reimers, Nils},
   TITLE      = {MTEB: Massive Text Embedding Benchmark},
   PUBLISHER  = {arXiv},
   JOURNAL    = {arXiv preprint arXiv:2210.07316},
   YEAR       = {2022},
   URL        = {https://arxiv.org/abs/2210.07316},
   DOI        = {10.48550/ARXIV.2210.07316},
}

@inproceedings{Brown:2020,
   AUTHOR     = {Brown, Tom
               AND Mann, Benjamin
               AND Ryder, Nick
               AND Subbiah, Melanie
               AND Kaplan, Jared D
               AND Dhariwal, Prafulla
               AND Neelakantan, Arvind
               AND Shyam, Pranav
               AND Sastry, Girish
               AND Askell, Amanda
               AND Agarwal, Sandhini
               AND Herbert-voss, Ariel
               AND Krueger, Gretchen
               AND Henighan, Tom
               AND Child, Rewon
               AND Ramesh, Aditya
               AND Ziegler, Daniel
               AND Wu, Jeffrey
               AND Winter, Clemens
               AND Hesse, Chris
               AND Chen, Mark
               AND Sigler, Eric
               AND Litwin, Mateusz
               AND Gray, Scott
               AND Chess, Benjamin
               AND Clark, Jack
               AND Berner, Christopher
               AND Mccandlish, Sam
               AND Radford, Alec
               AND Sutskever, Ilya
               AND Amodei, Dario},
   BOOKTITLE  = {Advances in Neural Information Processing Systems},
   EDITOR     = {H. Larochelle and M. Ranzato and R. Hadsell and M.F. Balcan and H. Lin},
   PAGES      = {1877--1901},
   PUBLISHER  = {Curran Associates, Inc.},
   TITLE      = {Language Models are Few-Shot Learners},
   URL        = {https://proceedings.neurips.cc/paper_files/paper/2020/file/1457c0d6bfcb4967418bfb8ac142f64a-Paper.pdf},
   VOLUME     = {33},
   YEAR       = {2020}
}

@inproceedings{Heckman:2008,
  	ADDRESS    = {Kaiserslautern Germany},
  	TITLE      = {On establishing a benchmark for evaluating static analysis alert prioritization and classification techniques},
  	ISBN       = {978-1-59593-971-5},
  	URL        = {https://dl.acm.org/doi/10.1145/1414004.1414013},
  	DOI        = {10.1145/1414004.1414013},
  	LANGUAGE   = {en},
  	URLDATE    = {2025-02-28},
  	BOOKTITLE  = {Proceedings of the {Second} {ACM}-{IEEE} international symposium on {Empirical} software engineering and measurement},
  	PUBLISHER  = {ACM},
  	AUTHOR     = {Heckman, Sarah AND Williams, Laurie},
  	MONTH      = oct,
  	YEAR       = {2008},
  	PAGES      = {41--50},
}

@article{Vaswani:2017,
   TITLE      = {Attention is all you need},
   AUTHOR     = {Vaswani, Ashish
               AND Shazeer, Noam
               AND Parmar, Niki
               AND Uszkoreit, Jakob
               AND Jones, Llion
               AND Gomez, Aidan N
               AND Kaiser, {\l}ukasz
               AND Polosukhin, Illia},
   JOURNAL    = {Advances in neural information processing systems},
   VOLUME     = {30},
   YEAR       = {2017}
}

@article{Wen:2024,
   AUTHOR     = {Wen, Cheng
               AND Cai, Yuandao
               AND Zhang, Bin
               AND Su, Jie
               AND Xu, Zhiwu
               AND Liu, Dugang
               AND Qin, Shengchao
               AND Ming, Zhong
               AND Cong, Tian},
   TITLE      = {Automatically Inspecting Thousands of Static Bug Warnings with Large Language Model: How Far Are We?},
   YEAR       = {2024},
   ISSUE_DATE = {August 2024},
   PUBLISHER  = {Association for Computing Machinery},
   ADDRESS    = {New York, NY, USA},
   VOLUME     = {18},
   NUMBER     = {7},
   ISSN       = {1556-4681},
   URL        = {https://doi.org/10.1145/3653718},
   DOI        = {10.1145/3653718},
   JOURNAL    = {ACM Trans. Knowl. Discov. Data},
   MONTH      = jun,
   ARTICLENO  = {168},
   NUMPAGES   = {34}
}

@article{Li:2024,
   AUTHOR     = {Li, Haonan
               AND Hao, Yu
               AND Zhai, Yizhuo
               AND Qian, Zhiyun},
   TITLE      = {Enhancing Static Analysis for Practical Bug Detection: An LLM-Integrated Approach},
   YEAR       = {2024},
   ISSUE_DATE = {April 2024},
   PUBLISHER  = {Association for Computing Machinery},
   ADDRESS    = {New York, NY, USA},
   VOLUME     = {8},
   NUMBER     = {OOPSLA1},
   URL        = {https://doi.org/10.1145/3649828},
   DOI        = {10.1145/3649828},
   JOURNAL    = {Proc. ACM Program. Lang.},
   MONTH      = apr,
   ARTICLENO  = {111},
   NUMPAGES   = {26}
}

@article{Katerina:2015,
   TITLE      = {On the capability of static code analysis to detect security vulnerabilities},
   JOURNAL    = {Information and Software Technology},
   VOLUME     = {68},
   PAGES      = {18-33},
   YEAR       = {2015},
   ISSN       = {0950-5849},
   DOI        = {https://doi.org/10.1016/j.infsof.2015.08.002},
   URL        = {https://www.sciencedirect.com/science/article/pii/S0950584915001366},
   AUTHOR     = {Katerina Goseva-popstojanova AND Andrei Perhinschi}
}

@article{Zou:2019,
   TITLE      = {$\mu$VulDeePecker: A Deep Learning-Based System for Multiclass Vulnerability Detection},
   ISSN       = {2160-9209},
   URL        = {http://dx.doi.org/10.1109/TDSC.2019.2942930},
   DOI        = {10.1109/tdsc.2019.2942930},
   JOURNAL    = {IEEE Transactions on Dependable and Secure Computing},
   PUBLISHER  = {Institute of Electrical and Electronics Engineers (IEEE)},
   AUTHOR     = {Zou, Deqing
               AND Wang, Sujuan
               AND Xu, Shouhuai
               AND Li, Zhen
               AND Jin, Hai},
   YEAR       = {2019},
   PAGES      = {1-1} }

@article{Li:2022,
   TITLE      = {SySeVR: A Framework for Using Deep Learning to Detect Software Vulnerabilities},
   VOLUME     = {19},
   ISSN       = {2160-9209},
   URL        = {http://dx.doi.org/10.1109/TDSC.2021.3051525},
   DOI        = {10.1109/tdsc.2021.3051525},
   NUMBER     = {4},
   JOURNAL    = {IEEE Transactions on Dependable and Secure Computing},
   PUBLISHER  = {Institute of Electrical and Electronics Engineers (IEEE)},
   AUTHOR     = {Li, Zhen
               AND Zou, Deqing
               AND Xu, Shouhuai
               AND Jin, Hai
               AND Zhu, Yawei
               AND Chen, Zhaoxuan},
   YEAR       = {2022},
   MONTH      = jul, pages={2244-2258} }

@inproceedings{Calcagno:2011,
   AUTHOR     = {Calcagno, Cristiano AND Distefano, Dino},
   TITLE      = {Infer: an automatic program verifier for memory safety of C programs},
   YEAR       = {2011},
   ISBN       = {9783642203978},
   PUBLISHER  = {Springer-Verlag},
   ADDRESS    = {Berlin, Heidelberg},
   BOOKTITLE  = {Proceedings of the Third International Conference on NASA Formal Methods},
   PAGES      = {459–465},
   NUMPAGES   = {7},
   LOCATION   = {Pasadena, CA},
   SERIES     = {NFM'11}
}

@misc{Wolf:2020,
      title={HuggingFace's Transformers: State-of-the-art Natural Language Processing}, 
      author={Thomas Wolf and Lysandre Debut and Victor Sanh and Julien Chaumond and Clement Delangue and Anthony Moi and Pierric Cistac and Tim Rault and Rémi Louf and Morgan Funtowicz and Joe Davison and Sam Shleifer and Patrick von Platen and Clara Ma and Yacine Jernite and Julien Plu and Canwen Xu and Teven Le Scao and Sylvain Gugger and Mariama Drame and Quentin Lhoest and Alexander M. Rush},
      year={2020},
      eprint={1910.03771},
      archivePrefix={arXiv},
      primaryClass={cs.CL},
      url={https://arxiv.org/abs/1910.03771}, 
}

@misc{Svyatkovskiy:2020,
      title={IntelliCode Compose: Code Generation Using Transformer}, 
      author={Alexey Svyatkovskiy and Shao Kun Deng and Shengyu Fu and Neel Sundaresan},
      year={2020},
      eprint={2005.08025},
      archivePrefix={arXiv},
      primaryClass={cs.CL},
      url={https://arxiv.org/abs/2005.08025}, 
}

@misc{sonarqube,
  author = {{SonarSource}},
  title = {{SonarQube: Continuous Code Quality and Security}},
  howpublished = {[\url{https://www.sonarsource.com/products/sonarqube/
      }]},
  year = {2025},
  note = {Accessed: 20 November 2025}
}

@misc{PMD,
  author = {{PMD}},
  title = {{An extensible cross-language static code analyzer.}},
  howpublished = {[\url{https://pmd.github.io/
      }]},
  year = {2025},
  note = {Accessed: 20 November 2025}
}

@misc{SpotBugs,
  author = {{SpotBugs}},
  title = {{Find bugs in Java Programs}},
  howpublished = {[\url{https://spotbugs.github.io/
      }]},
  year = {2025},
  note = {Accessed: 20 November 2025}
}

@Article{Koszo:2025,
author={K{\'o}sz{\'o}, D{\'a}vid
and Aladics, Tam{\'a}s
and Ferenc, Rudolf
and Heged{\H{u}}s, P{\'e}ter},
title={A Large-Scale Collection Of (Non-)Actionable Static Code Analysis Reports},
journal={Scientific Data},
year={2025},
month={Nov},
day={28},
volume={12},
number={1},
pages={1884},
issn={2052-4463},
doi={10.1038/s41597-025-06154-7},
url={https://doi.org/10.1038/s41597-025-06154-7}
}

\end{document}